**Comparing optical oscillators across the air to milliradians in phase and $10^{-17}$ in frequency**

Laura C. Sinclair[1], Hugo Bergeron[2], William C. Swann[1], Esther Baumann[1], Jean-Daniel Deschenes[2], and Nathan R. Newbury[1]

[1]*National Institute of Standards and Technology, 325 Broadway, Boulder, Colorado 80305*
[2]*Université Laval, 2325 Rue de l'Université, Québec, QC, G1V 0A6, Canada*

**Abstract**

We demonstrate carrier-phase optical two-way time-frequency transfer (carrier-phase OTWTFT) through the two-way exchange of frequency comb pulses. Carrier-phase OTWTFT achieves frequency comparisons with a residual instability of $1.2 \times 10^{-17}$ at 1 second across a turbulent 4-km free space link, surpassing previous OTWTFT by 10-20x and enabling future high-precision optical clock networks. Furthermore, by exploiting the carrier-phase, this approach is able to continuously track changes in the relative optical phase of distant optical oscillators to 9 mrad (7 attoseconds) at 1-sec averaging, effectively extending optical phase coherence over a broad spatial network for applications such as correlated spectroscopy between distant atomic clocks.

Applications of future optical clock networks include time dissemination, chronometric geodesy, coherent sensing, tests of relativity, and searches for dark matter among others [1–14]. This promise has motivated continued advances in optical clocks and oscillators [15–19] and in the optical transfer techniques to network them. In particular, time-frequency transfer over fiber-optic networks has seen tremendous progress [1,7,20–23]. However, many applications require clock networks connected via free-space links. Direct adoption of fiber-based approaches to free-space is possible but hampered by atmospheric turbulence [24]. Satellite-laser-ranging approaches such as T2L2 are being actively explored [25–28]. Here, we consider optical two-way time-frequency transfer (OTWTFT) based on the two-way exchange of frequency comb pulses [24,29–35]. This approach exploits the reciprocity (equality) in the time-of-flight for light to travel each direction across a single-mode link [36], just as in rf-based two-way satellite time-frequency transfer [37–39] and analogous fiber-optic demonstrations [23,40–42]. In previous work, this OTWTFT approach used the arrival time of the frequency comb pulses to support frequency comparisons at residual instabilities of $\sim 4\times 10^{-16}$ at 1-second averaging times [29], and ultimately to enable sub-femtosecond time synchronization of distant optical and microwave-based clocks [32–34].

Here, we demonstrate OTWTFT can exploit the carrier phase of the frequency comb pulses for much higher performance. While carrier-phase measurements are relatively straightforward across optical fiber because of the uninterrupted stable signal, the same is not true of a free-space link where atmospheric turbulence leads to strong phase noise and signal intermittency, in turn presenting a severe challenge to "unwrapping" the measured phase without catastrophic $\pm\pi$ phase errors. We show such phase unwrapping is possible over hour durations, despite atmospheric turbulence, and despite strong phase drift between the distant sites. We achieve frequency comparison with a residual instability (modified Allan deviation) of $1.2\times10^{-17}$ at 1 second: 10-20 times lower than achieved with the pulse timing alone. This instability drops to $6\times10^{-20}$ at 850 s. We show the short-term residual instability is near the limit given by atmospheric-turbulence-driven reciprocity breakdown. Most importantly, it is well below the absolute instability of even the best optical clocks and oscillators.

Carrier-phase OTWTFT essentially tracks the evolution of the relative optical phase between the two distant optical oscillators. Specifically, here we track the ~300 million-radian residual phase evolution between our two 1535-nm cavity-stabilized lasers without ambiguity to within a 0.2-rad standard deviation at 400-μs time resolution. The corresponding time deviation reaches 7 attoseconds (9 mrad) at 1-s averaging time. The relative phase noise power spectral density drops below $10^{-4}$ rad$^2$/Hz (~ 60 as$^2$/Hz) at 1 Hz offset, or >25 dB below that achievable with pulse timing alone. In this sense, we establish tight mutual optical phase coherence between sites that could be exploited in future applications requiring spatially distributed phase coherence. In particular, several groups have compared optical atomic clocks to ultra-high precision by cancelling out common-mode optical phase noise of the clocks' local oscillators. Takamoto *et al.* demonstrated synchronous sampling of two distant atomic ensembles, avoiding the Dick effect [43], while Chou *et al.* and others demonstrated correlated spectroscopy to extend the Ramsey interrogation times beyond the local oscillator coherence time [44–46]. Carrier-phase OTWTFT could enable such distant optically coherent measurements even with portable (but noiser) cavity-stabilized lasers [47–50] and even over turbulent links.

To successfully track the optical phase, the interval between phase measurements must be shorter than the mutual phase coherence time between the distant optical oscillators. (For the same reason an optical clock's Ramsey interrogation time is limited by the local oscillator coherence time.) Otherwise, the resulting $\pm\pi$ phase ambiguities lead to complete loss of frequency/phase information. This presents two problems for carrier-phase OTWTFT over a turbulent atmosphere. First, atmospheric turbulence scrambles the received light's optical phase, degrading the mutual optical coherence time. This problem is circumvented by exploiting the time-of-flight reciprocity. Second, atmospheric turbulence causes fades (signal loss) at random times, with random durations and random separations, so the measurement interval often exceeds the mutual coherence time (of 50 ms here). This problem is circumvented by combining the timing information from both the pulse's carrier-phase and envelope to extend the coherence across the fades.

We use a folded 4-km link (Fig. 1) to compare optical oscillators at site A and B. At each site, a cavity-stabilized laser serves as the optical oscillator. We let the oscillator at site A define the timescale with known frequency $\nu_A$ and phase $\varphi_A(t) = 2\pi\nu_A t$. The optical phase of site B's oscillator is $\varphi_B(t) = 2\pi\nu_B t + \delta\varphi(t)$, where the first term is the phase evolution from the *a priori* estimated frequency $\nu_B \neq \nu_A$ and the second captures the unknown phase wander or equivalently timing wander $\delta\tau = (2\pi\nu_B)^{-1}\delta\varphi$. The unknown frequency variation is $\delta\nu(t) = (2\pi)^{-1} d\delta\varphi(t)/dt$. Our objective is to measure $\delta\varphi(t)$ or equivalently $\delta\nu(t)$. We drop any constant phase offset by setting $\delta\varphi(0) = 0$ (and so $\delta\tau(0) = 0$) since its knowledge requires full time synchronization. The folded link permits truth data acquisition via a direct optical heterodyne measurement, which also necessitates both oscillators operate at nearly the same frequency, here at $\nu_A$ = 194.584000 THz and $\nu_B$ = 194.584197 THz. In general, however, the optical oscillators would be at widely different frequencies and locations. Additional details on the combs and free-space terminals are in Refs. [32,51] with additional experimental values given in Supplemental Table 1.

At each site, the optical oscillator signal is transferred via a Doppler-cancelled fiber link to a self-referenced frequency comb where its phase is mapped onto the comb. Specifically, at site B the relative phase noise $\delta\varphi(t)$ maps to noise in both the optical phase and the timing of comb B's pulse train. This noise is not white but includes strong random frequency walk. We compare the phase and timing of comb B with comb A via linear optical sampling in a two-way configuration. To do this, combs A and B are phase locked using our *a priori* information of $\nu_A$ and $\nu_B$ such that their repetition frequencies $f_{r,A}$ and $f_{r,B}$ differ by $\Delta f_r = f_{r,B} - f_{r,A}$. Here, $f_{r,A}$ = 200 MHz with $\Delta f_r \approx$ 2.46 kHz. At each site, we filter the comb to a ~1-THz bandwidth around 1560 nm and transmit it to the opposite site where it is heterodyned against the local comb to generate a series of cross-correlations, which are analyzed to extract $\delta\varphi(t)$. Note the transmitted comb optical spectrum need not – and does not – encompass the optical oscillator frequencies.

The extraction of $\delta\varphi(t)$ proceeds as follows. For convenience, we lock the self-referenced combs such that $f_{r,A} = \nu_A/n_A$ and $f_{r,B} = \nu_B/n_B$ where $n_A$ and $n_B$ are the indices of the comb tooth nearest to the local oscillator at sites A and B. We then identify the pair of comb tooth frequencies, $\tilde{\nu}_A$ and $\tilde{\nu}_B$, nearest to the center of the transmitted optical spectrum having a frequency separation $\Delta\tilde{\nu} \equiv \tilde{\nu}_B - \tilde{\nu}_A < \pm\Delta f_r/2$. This pair, rather than $\nu_A$ and $\nu_B$ directly, will serve as the carrier frequencies for the carrier-phase OTWTFT, as shown below. At site A, we write the transmitted and received comb electric fields with respect to this pair as

$$E_A(t) = e^{i2\pi\tilde{\nu}_A t} \sum_m E_{A,m} e^{i2\pi m f_{r,A} t}$$
$$E_B(t) = e^{i2\pi\tilde{\nu}_B(t-T_{link})} e^{i\tilde{\nu}_B \delta\varphi/\nu_B} \sum_m E_{B,m} e^{i2\pi m f_{r,B}(t+\delta\tau-T_{link})} \quad (1)$$

where $E_{X,m}$ is the electric field, $m$ is the comb index from the tooth at $\tilde{\nu}_X$, $T_{link}$ is the slowly varying time-of-flight, and $\delta\tau = (2\pi\nu_B)^{-1}\delta\varphi$ is the timing jitter of comb B. The equations for site B are analogous, except that $T_{link}$ appears in $E_A(t)$. Note the unknown phase wander of oscillator B appears both in the timing noise, $\delta\tau$, and in the carrier optical phase of comb B. At each site, the combs are heterodyned to give a series of cross-correlations with complex envelopes $I_X(t)$, labelled by the integer $p$

$$V_X(t) \propto e^{i\Theta_{p,X}} \sum_p I_X(t - t_{p,X}), \quad (2)$$

assuming $T_{link}$ and $\delta\tau$ vary slowly on the timescales of $1/\Delta f_r$. (See Supplemental material for derivation.) The cross-correlation envelope peaks at times

$$t_{p,A} = \Delta f_r^{-1} \{p + f_{r,B} T_{link} - f_{r,B} \delta\tau\}$$
$$t_{p,B} = \Delta f_r^{-1} \{p - f_{r,A} T_{link} - f_{r,B} \delta\tau\} \quad (3)$$

with phase,

$$\Theta_{p,A} = 2\pi\Delta\tilde{\nu} t_{p,A} - 2\pi\tilde{\nu}_B T_{link} + \tilde{\nu}_B \nu_B^{-1} \delta\varphi$$
$$\Theta_{p,B} = 2\pi\Delta\tilde{\nu} t_{p,B} + 2\pi\tilde{\nu}_A T_{link} + \tilde{\nu}_B \nu_B^{-1} \delta\varphi. \quad (4)$$

As expected for a two-way measurement, the time-of-flight enters with an opposite sign at the two sites in (3) and (4) and can thus be eliminated. Note the cross-correlations do not occur simultaneously at the two sites, rather asynchronously with offset $|t_{p,A} - t_{p,B}| < 2/\Delta f_r$, which will be important later. For each site, we evaluate $t_{p,X}$ and $\Theta_{p,X}$ at $1/\Delta f_r \sim$ 400 μs intervals via matched filter processing against the $p=0$ cross-correlation thereby dropping any overall constant time/phase offsets.

In previous OTWTFT, we effectively solved (3) for $\delta\tau(t_p)$ evaluated at $t_p \equiv (t_{p,A} + t_{p,B})/2$, from which we extracted the fractional frequency uncertainty $\delta\nu/\nu_B = d\delta\tau/dt$. However, the precision of $\delta\tau(t_p)$ is typically SNR-limited to 3 to 8 fs (4 to 10 radians equivalent optical-phase uncertainty).

In carrier-phase OTWTFT, we exploit the cross-correlation phase for higher precision by solving (4) to find

$$\delta\varphi(t_p) \approx \frac{\nu_B}{2\tilde{\nu}_B}\left\{\Theta_{p,A} + \Theta_{p,B} - 4\pi\Delta\tilde{\nu}t_p + 2\pi\Delta\tilde{\nu}T_{link} - 2\pi k_p\right\}, \tag{5}$$

dropping the next term $\pi\nu_B(t_{p,A} - t_{p,B})dT_{link}/dt$. (See Supplemental material.) After determining $T_{link}$ from (3), all the terms are known except for $k_p$, which is a time-dependent integer accounting for the π-ambiguity (~2.5 fs equivalent timing uncertainty) in this phase measurement. The precision is now limited by the ~0.1 radian noise typical of the comb phase locks and Doppler-cancelled links for a total uncertainty of ~0.2 radians, corresponding to 160 attoseconds in timing precision at the 400-μs update rate.

Of course, this higher precision is lost in π-ambiguities unless $k_p$ is known. If $\delta\varphi$ varies slowly with successive measurements, standard unwrapping algorithms can track $k_p$. However, $\delta\varphi$ varies significantly from mutual phase noise between the oscillators, characterized by the measured power spectral density (PSD) of $S_{\delta\varphi} = 22f^{-4}$ rad²/Hz, where $f$ is the Fourier frequency. More importantly, random fades from turbulence-induced scintillation, physical obstructions, or loss of terminal pointing cause measurement gaps well beyond $1/\Delta f_r$ ~ 400 µs. Therefore, a layered Kalman-filter-based unwrapping algorithm is used. (See Supplemental material.) The inputs are the first four terms of (5), $S_{\delta\varphi}$, $\delta\tau(t_p)$ from (3), the received power, and the power-dependent uncertainty in $t_{p,X}$ and $\Theta_{p,X}$. The output is a prediction of the phase, which is compared with the observed phase to find $k_p$. The Kalman filter also predicts the uncertainty $\sigma_{\varphi,p}$ in the predicted phase which grows with time over long fades, eventually leading to ambiguity in $k_p$ and requiring use of the envelope timing to re-acquire $k_p$. Indeed, a functional, rigorous definition of mutual coherence time is exactly the time interval until the predicted phase's uncertainty exceeds a value $\sigma_\varphi$, denoted $t_{coh}^{\sigma_\varphi}$. (This coherence time differs from frequency-domain definitions based on linewidth or PSDs which are poorly defined for $S_{\delta\varphi} \propto f^{-4}$, and is in fact closely related to the relevant coherence for Ramsey interrogation [19]). For our system, $t_{coh}^{1\,\text{rad}}$ ~50 ms. However, the algorithm uses a stricter limit of $t_{coh}^{0.12\,\text{rad}}$ ~ 7 ms before reverting to the envelope timing to "re-acquire" $k_p$. While all processing is currently offline, real-time processing following Ref. [32] is possible.

Figure 2a shows the resulting unwrapped phase $\delta\varphi(t)$ over a ~1.4-hour measurement across the 4-km turbulent link. It is dominated by a roughly linear frequency drift, leading to over 300 million radians of total phase drift (beyond the expected phase drift of $2\pi(\nu_B - \nu_A)t$) with random phase wander reflecting the $f^{-4}$ PSD, as shown in the inset. Therefore, phase continuity of the measured $\delta\varphi(t)$ can only be evaluated by comparison with truth data, $\delta\varphi_{truth}(t)$, acquired from the direct shorted heterodyne beat between oscillators. As shown in Fig. 2b, $\delta\varphi(t) - \delta\varphi_{truth}(t)$ shows no phase slips. The standard deviation is 0.2 rad (or 160 attoseconds in time units) at the full 400-μs sample rate and 30 mrad (24 attoseconds) at a 1-s time resolution. Finally, a linear fit to $\delta\varphi(t) - \delta\varphi_{truth}(t)$ yields the overall accuracy in the determination of oscillator B's frequency offset $\delta v$ across the measurement, which is 2 μHz, corresponding to a fractional uncertainty of $10^{-20}$.

Although it is not evident in the densely plotted data of Fig. 2a-b, fades occur during 1% of the total 1.4 hours. Because of turbulence, for ~3 mW transmitted power, the received power varied from 0 to 5 μW with a detection threshold of 10 nW, below which a fade (signal loss) occurs. Fades with durations beyond $t_{coh}^{0.12\,rad}$ ~ 7 ms require re-acquisition of the phase via the envelope. Figure 2c-d show examples of phase-continuous measurement across a single fade and across multiple juxtaposed fades. For the data of Fig. 2b, there are ~1400 fades randomly distributed in time with durations beyond $t_{coh}^{0.12\,rad}$ ~ 7 ms, while a later run had 26% fades with ~28,500 fades beyond $t_{coh}^{0.12\,rad}$. (See Supplemental Figure 1.)

Figure 3 shows the phase noise PSD for $\delta\varphi(t) - \delta\varphi_{truth}(t)$ of Fig. 2b, and compares this PSD to previous OTWTFT using the envelope only (i.e. finding $\delta\tau$ from (3) only). Above 1 Hz, the carrier-phase data is > 25 dB lower, with a floor of ~ $3\times10^{-5}$ rad$^2$/Hz (~ 20 as$^2$/Hz). Below 40 mHz, the two PSDs converge as the noise is limited by flicker (1/$f$) noise from variations in the delays within the transceivers.

Figure 4 shows the modified Allan deviation from $\delta\varphi(t) - \delta\varphi_{truth}(t)$ at both 1% fades (e.g. Fig. 2b) and 26% fades. At 1% fades, the carrier-phase OTWTFT instability is $1.2\times10^{-17}/t_{avg}^{3/2}$ from 0.01 s to a few seconds. It then flattens from a few seconds to 10 s likely due to fluctuations in the transceiver delays from air-conditioning cycling, before dropping to $6\times10^{-20}$ at 850 s. At short times, the carrier-phase OTWTFT is 10 times lower than for the envelope-only approach of previous OTWTFT. At the higher fade rate of 26%, the carrier-phase OTWTFT rises to $5.6\times10^{-17}$ at 1 s and the envelope-only OTWTFT is 20 times higher still.

The measured instability of $\sim 10^{-17}$ at 1 second translates to a time deviation of 7 attoseconds, or equivalently 9 mrad, at one second, indicating reciprocity for a single-mode link holds to a remarkable degree even across a turbulent atmosphere. Nevertheless, there is a slight discrepancy between the open-path Allan deviation of $1.2\times10^{-17}$ and shorted (no open-path) Allan deviation of $0.95\times10^{-17}$. (See Fig. 4.) We attribute this discrepancy to a slight breakdown in reciprocity from asynchronous sampling and time-dependent turbulence, i.e. exactly the additional term $\propto 0.5(t_{p,A}-t_{p,B})dT_{link}/dt$ discussed after Eq. (5). (Other effects that limit reciprocity [30,31,52] are unobserved in this configuration.) The time-dependent atmospheric piston phase noise, i.e. $T_{link}(t)$, is characterized by a spectral noise density of $af^{-7/3}$ [24,53] where $f$ is the Fourier frequency and $a \sim 10^{-28} \text{s}^2\text{Hz}^{4/3}$. Approximating this PSD as $\sim f^{-2}$ gives a contribution to the modified Allan deviation of

$$\sigma_{atm}(t_{avg}) = \pi\sqrt{3a/2}\left|t_{p,A}-t_{p,B}\right|t_{avg}^{-3/2} .\qquad(6)$$

The asynchronous sampling, $\left|t_{p,A}-t_{p,B}\right|$, ranges from 0 to $(2\Delta f_r)^{-1} \sim 200\,\mu s$; the shaded region in Fig. 4 shows $\sigma_{atm}(t_{avg})$ for a 10-90% range (20 µs< $\left|t_{p,A}-t_{p,B}\right|$ < 180 µs). The quadrature sum of the shorted Allan deviation and $\sigma_{atm}(t_{avg})$ at $|t_{p,A} - t_{p,B}| = 100$ $\mu$s agrees with the measured open-path results of $1.2\times10^{-17}$ at 1 second.

The 4-km link distance demonstrated here is not the maximum range limit. Indeed, carrier-phase OTWTFT poses no additional constraints on range, excepting that the power-aperture product must be increased along with the link distance to maintain sufficient received comb power, comparable to the power requirements for coherent communications. At link distances >60 km, the time-of-flight reaches $(2\Delta f_r)^{-1}$ and the timestamps must be properly aligned to avoid an increase in the asynchronous sampling noise floor given by Eqn. (6).

We have demonstrated phase comparisons between optical oscillators or clocks using the carrier phase of frequency comb pulses over turbulent free-space paths. Carrier-phase OTWTFT reaches $1.2\times10^{-17}$ fractional stability at 1 second averaging time, corresponding to a time deviation of 7 attoseconds, despite the presence of turbulence-induced fades. In so doing, it connects the optical phases as distant sites and should enable correlated spectroscopy of distant optical clocks.

**Acknowledgments**
We thank Danielle Nicolodi, Kevin Cossel and Fabrizio Giorgetta for helpful discussions and technical assistance from Michael Cermak and Isaac Khader. We acknowledge funding from the DARPA DSO PULSE program and NIST.

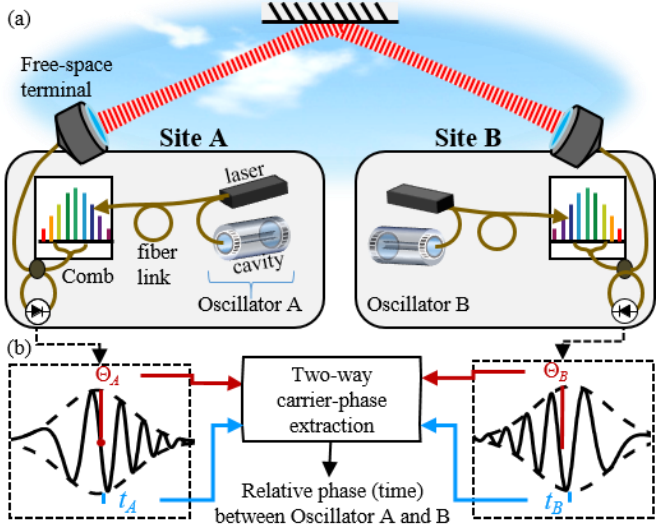

Figure 1: (a) Experimental setup. The phase of the local optical oscillator (cavity-stabilized laser) is transferred via a Doppler-cancelled fiber link to a frequency comb, a portion of whose output is transmitted to the opposite site, where it is heterodyned against the local comb. (b) The resulting cross-correlation between pulse trains is analysed to extract the envelope peak time, $t_{p,X}$, and the phase, $\Theta_{p,X}$, which are input to a Kalman-filter based algorithm to calculate the relative phase/timing evolution between the two optical oscillators despite atmospheric phase noise and fading.

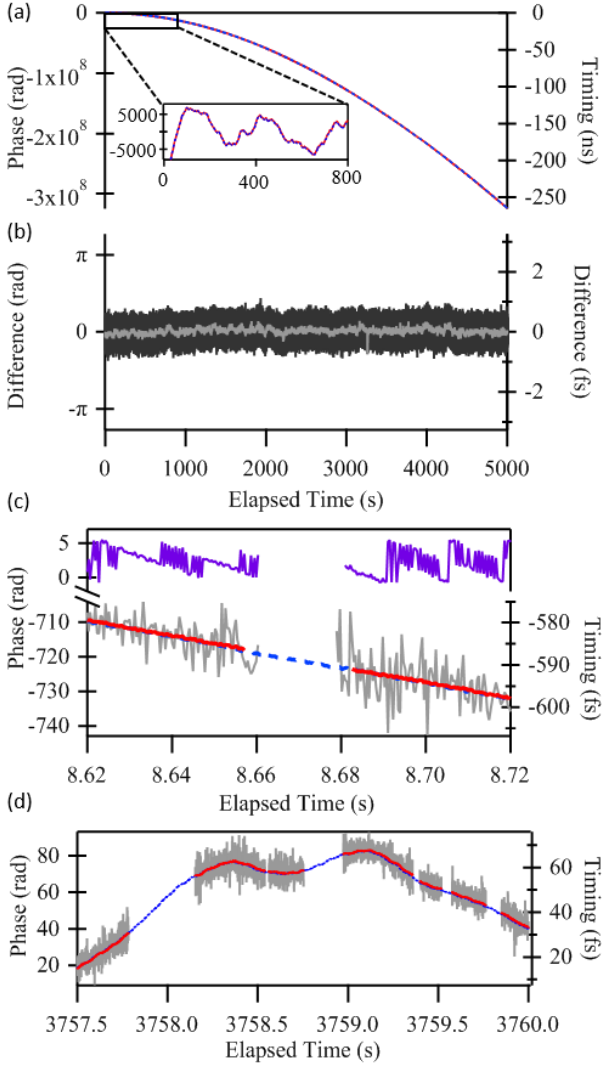

Figure 2: Results for ~1.4-hours across the turbulent 4-km link. (a) Oscillator B's residual phase from carrier-phase OTWTFT, $\delta\varphi(t)$, (red line) and from direct oscillator-to-oscillator truth data, $\delta\varphi_{truth}(t)$, (dashed blue line) in radians (left axis) and scaled to time units by $(2\pi\nu_B)^{-1}$ (right axis). For both, we set the value at $t=0$ to zero; consequently, we use "timing" on the right axis, emphasizing that the overall time offset between sites is unknown. The dominant quadratic behaviour arises from the ~4 Hz/s frequency drift between the optical oscillators. Inset: phase wander after removing a quadratic fit illustrating the phase fluctuations at all time scales expected from the $1/f^4$ relative phase noise. (b) Difference between the carrier-phase OTWTFT and truth data, $\delta\varphi(t) - \delta\varphi_{truth}(t)$, at 400 µsec sampling (black) with 0.2 rad (160 as) standard deviation and at 1-second averaging (gray) with 30 mrad (24 as) standard deviation. There are no phase discontinuities over the entire period. The average slope yields an overall frequency difference between truth data and the OTWTFT data of 2 µHz, despite the accumulated 18 kHz offset between the oscillators. (c) A 10-second segment showing the phase before unwrapping (top panel, purple) and after (red line), which follows the truth data. Also shown is the envelope timing (gray line),

used to unwrap the phase across fades. (d) Similar to (c) but illustrating phase continuity over a complicated fade sequence. (An overall slope of 40 rad/sec was removed for display purposes.)

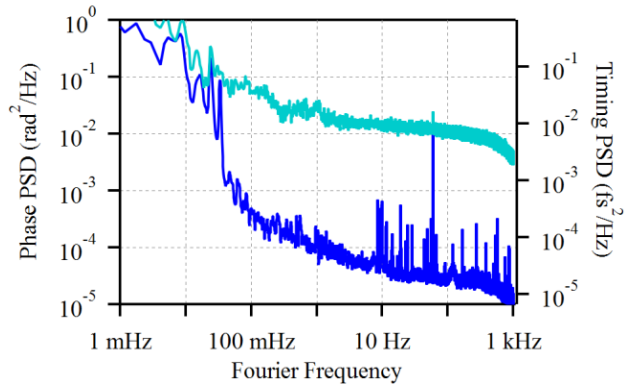

Figure 3: Phase noise power spectral density of $\delta\varphi(t) - \delta\varphi_{truth}(t)$ (dark blue) in rad²/Hz (left axis) and converted to fs²/Hz (right axis). For comparison, the corresponding power spectral density extracted from the envelope pulse timing alone is also shown (light blue).

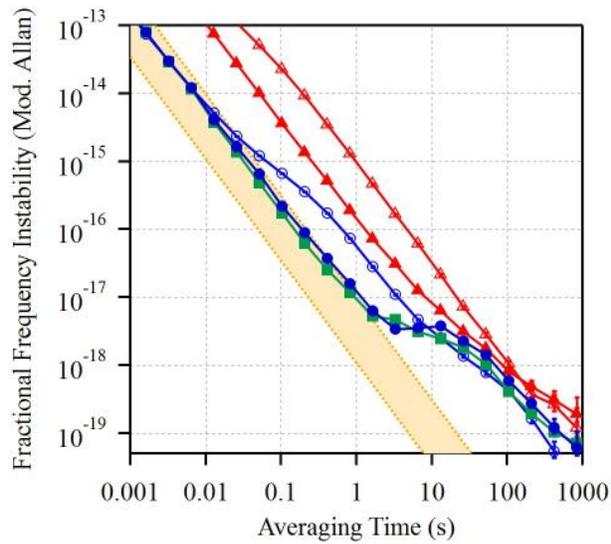

Figure 4: Residual fractional frequency instability (modified Allan deviation) for carrier-phase OTWTFT over a 4-km link with 1% fades (blue circles) and 26% fades (open red circles) compared to the corresponding envelope-only OTWTFT for 1% fades (blue triangles) and 26% fades (open red triangle). The carrier-phase OTWTFT instability over a shorted (0 km) link is also shown (green squares). Finally, the fundamental limit set by the time-dependence of the atmospheric turbulence is indicated shaded orange box (at 10-90% likelihood).

Supplemental Material:

1. Supplemental Material: Detailed derivation of Eqs. (2) - (5)

The notation here follows the main manuscript and so definitions are not repeated. Frequency combs are usually defined in terms of the carrier-envelope offset frequency, $f_{ceo}$, and repetition rate, $f_r$, so that the comb tooth frequencies are given by $\nu_n = f_{ceo} + n f_r$, where $n$ is an integer. We can alternatively write the comb tooth frequencies as $\nu_m = \tilde{\nu} + m f_r$, where $\tilde{\nu} \equiv f_{ceo} + \tilde{n} f_r$ is a specific comb tooth frequency at $n = \tilde{n}$ and $m$ represents an integer offset from this tooth. Thus, combs A and B at their respective sites located at $z = z_A$ and $z = z_B$ become

$$E_A(t, z_A) = e^{i2\pi\tilde{\nu}_A t} \sum_m E_{A,m} e^{i2\pi m f_{r,A} t} \qquad (7)$$

$$E_B(t, z_B) = e^{i2\pi\tilde{\nu}_B t + i\tilde{\nu}_B \delta\varphi/\nu_B} \sum_m E_{B,m} e^{i2\pi m f_{r,B}(t+\delta\tau)} \qquad (8)$$

where $\delta\tau \equiv (2\pi\nu_B)^{-1} \delta\varphi$, dropping any overall phase or time offsets and without yet selecting $\tilde{n}_A$ or $\tilde{n}_B$. Here, as in the main text, we assign all the phase noise to Site B which then appears with appropriate scaling on Comb B in Eq. (8). At the opposite site, each comb will have travelled across distance $|z_A - z_B|$ with phase velocity $v_{phase}$ and group velocity $v_{group}$. Ignoring higher order dispersion, the comb fields at the opposite site are then

$$E_A(t, z_B) = e^{i2\pi\tilde{\nu}_A(t-T_{link})} \sum_m E_{A,m} e^{i2\pi m f_{r,A}(t - T_{link}^{group})} \qquad (9)$$

$$E_B(t, z_A) = e^{i2\pi\tilde{\nu}_B(t-T_{link}) + i\tilde{\nu}_B \delta\varphi/\nu_B} \sum_m E_{B,m} e^{i2\pi m f_{r,B}(t+\delta\tau - T_{link}^{group})} \qquad (10)$$

where $T_{link} = |z_A - z_B|/v_{phase}$ is the time-of-flight associated with the phase and $T_{link}^{group} = |z_A - z_B|/v_{group}$ is the slightly different time-of-flight associated with pulse envelope. As the extracted timings will be subtracted off between the two directions before further processing

(and conversely for the carrier phase), any change in either $T_{link}$ or $T_{link}^{group}$ that is reciprocal will cancel. Any non-reciprocal path through dispersive media (fiber in the transceivers for example) will, however, change these group and phase delays differently and thus there exists a potential offset. To deal with this offset, the phase tracking algorithm applies an offset when re-acquiring $\delta\varphi$ by use of the scaled $\delta\tau$ from the pulse envelope timing, thereby absorbing any additional fixed or slowly-varying offset, even from non-reciprocal propagation. We will thus assume $T_{link}^{group} \approx T_{link}$ for the rest of the derivation.

With $T_{link}^{group} = T_{link}$, we recognize Eqs. (7) and (10) as exactly Eq. (1) in the main text, and Eqs. (8) and (9) as the analogous equations at the other site. At site A, the comb fields are overlapped on a detector to yield

$$E_B(t, z_A) E_A^*(t, z_A) = e^{i2\pi\Delta\tilde{\nu}t} e^{-i2\pi\tilde{\nu}_B T_{link} + i\tilde{\nu}_B \delta\varphi/\nu_B} \sum_{m,m'} E_{B,m} E_{A,m'}^* e^{-i2\pi m' f_{r,A} t} e^{i2\pi m f_{r,B}(t+\delta\tau - T_{link})}$$

(11)

where $\Delta\tilde{\nu} \equiv \tilde{\nu}_B - \tilde{\nu}_A$ and with a similar overlap at site B. This heterodyne field will be detected and sampled at $f_{r,A}$. We now impose the following three conditions, which are satisfied in the experiment. First, the optical bandwidth (i.e. non-zero amplitude comb teeth) spans less than $f_{r,A} f_{r,B}/(2\Delta f_r)$. Second, the heterodyne spectrum does not cross zero (e.g. for all non-zero amplitude comb teeth the frequency of comb A is above comb B, or vice versa). Third, we low-pass filter the detected signal to below $f_r/2$. Then, if we define the carrier frequencies (i.e. values of $\tilde{n}_A$ or $\tilde{n}_B$) such that $|\Delta\tilde{\nu}| < \Delta f_r/2$, it must be that $m = m'$ in the double sum and the detected signal at Site A and B is

$$V_A(t) = e^{i\Theta_A(t)} \sum_m I_{A,m} e^{i2\pi m \Delta f_r \left\{ t - \left( f_{r,B}/\Delta f_r \right) T_{link} + \left( f_{r,B}/\Delta f_r \right) \delta\tau \right\}} \tag{12}$$

$$V_B(t) = e^{i\Theta_B(t)} \sum_m I_{B,m} e^{i2\pi m \Delta f_r \left\{ t + \left( f_{r,A}/\Delta f_r \right) T_{link} + \left( f_{r,B}/\Delta f_r \right) \delta\tau \right\}}, \tag{13}$$

where the phases are $\Theta_A(t) = 2\pi \Delta \tilde{\nu} t - 2\pi \tilde{\nu}_B T_{link} + \tilde{\nu}_B \tilde{\nu}_B^{-1} \delta\varphi$ and $\Theta_B(t) = 2\pi \Delta \tilde{\nu} t + 2\pi \tilde{\nu}_A T_{link} + \tilde{\nu}_B \tilde{\nu}_B^{-1} \delta\varphi$, the complex amplitudes are $I_{A,m} \equiv E_{B,m} E_{A,m}^* R_{A,m}$ and $I_{B,m} \equiv E_{A,m} E_{B,m}^* R_{B,m}$, and $R_{X,m}$ accounts for the low-pass detector response at $X = A$ or $B$. The voltage at site A is sampled at discrete times $t_k = k f_{r,A}^{-1}$ and at site B at discrete times $t_k = k \left( f_{r,B} + \delta f_r \right)^{-1}$ where $k$ is an integer. (The system digitizes the real part of the cross-correlation, but a Hilbert transform is used to generate the analytical cross-correlation written here.)

The sums are over rf tones at frequencies $m \Delta f_r$. The sum of these tones is just a series of repeated cross-correlations or interferograms as the comb pulse trains walk across each other in time. The envelope of the successive cross-correlations is a maximum whenever the exponent in the sums of Eq. (12) or (13) is an integer, $p$. This integer then acts as a label for successive cross-correlations. For example, in Eq. (12), the $p^{th}$ cross-correlation is centered at the time $t_{p,A}$ that satisfies $\Delta f_r t_{p,A} - f_{r,B} T_{link} + f_{r,B} \delta\tau = p$ or $t_{p,A} = \Delta f_r^{-1} \left\{ p + f_{r,B} T_{link} - f_{r,B} \delta\tau \right\}$. The envelope of each cross-correlation is identical, given by the Fourier Transform of the $I_{A,m}$ and written here as $I_A(t)$, with a similar notation at Site B (assuming fixed dispersion across the link) Following this, we rewrite (12) and its counterpart (13) in a more useful form as the series of cross-correlations,

$$V_A(t) = e^{i\Theta_A(t)} \sum_p I_A \left( t - t_{p,A} \right) \tag{14}$$

$$V_B(t) = e^{i\Theta_B(t)} \sum_p I_B(t - t_{p,B}) \tag{15}$$

where $t_{p,B} = \Delta f_r^{-1} \{p - f_{r,A} T_{link} - f_{r,B} \delta\tau\}$. If we re-define $I_A(t) \to e^{i2\pi\Delta\tilde{v}t} I_A(t)$, then the measured voltages are

$$V_A(t) = e^{i\Theta_{p,A}} \sum_p I_A(t - t_{p,A}) \tag{16}$$

$$V_B(t) = e^{i\Theta_{p,B}} \sum_p I_B(t - t_{p,B}) \tag{17}$$

With $\Theta_{p,A} = 2\pi\Delta\tilde{v} t_{p,A} - 2\pi\tilde{v}_B T_{link} + \tilde{v}_B v_B^{-1} \delta\varphi$ and $\Theta_{p,B}(t) = 2\pi\Delta\tilde{v} t_{p,B} + 2\pi\tilde{v}_A T_{link} + \tilde{v}_B v_B^{-1} \delta\varphi$, which is Eq. (2) of the main text.

The matched filter processing at the two sites extracts

$$\left[\Theta_{p,A}\right]_{MF} = \Theta_{p,A} + 2\pi k_{p,A} \tag{18}$$

$$\left[\Theta_{p,B}\right]_{MF} = \Theta_{p,B} + 2\pi k_{p,B} \tag{19}$$

whose sum is

$$\left[\Theta_{p,A}\right]_{MF} + \left[\Theta_{p,B}\right]_{MF} = 2\pi\Delta\tilde{v}\left(t_{p,A} + t_{p,B}\right) + \tilde{v}_B v_B^{-1}\left[\delta\varphi(t_{p,A}) + \delta\varphi(t_{p,B})\right] \\ + 2\pi\left[\tilde{v}_A T_{link}(t_{p,B}) - \tilde{v}_B T_{link}(t_{p,A})\right] + 2\pi k_p \tag{20}$$

where $k_p \equiv k_{p,A} + k_{p,B}$ is an integer and we now explicitly include the time-dependence of the phase noise as $\delta\varphi(t)$ and of the time-of-flight as $T_{link}(t)$. Assuming both vary slowly with respect to $\Delta t_p \equiv (t_{p,B} - t_{p,A})/2$, a Taylor series expansion about $t_p \equiv (t_{p,A} + t_{p,B})/2$ yields

$$\begin{aligned}
\left[\Theta_{p,A}\right]_{MF} + \left[\Theta_{p,B}\right]_{MF} &= 4\pi\Delta\tilde{v}t_p + 2\tilde{v}_B v_B^{-1}\delta\varphi(t_p) - 2\pi\Delta\tilde{v}T_{link}(t_p) + 2\pi k_p \\
&+ 4\pi\tilde{v}_B\Delta t_p\dot{T}_{link}(t_p) - 2\pi\Delta\tilde{v}\Delta t_p\dot{T}_{link}(t_p) + \tilde{v}_B v_B^{-1}\Delta t_p^2\delta\ddot{\varphi}(t_p)
\end{aligned} \quad (21)$$

The first line contains the dominant term. It is inverted to solve for $\delta\varphi(t_p)$ as

$$\delta\varphi(t_p) \approx \frac{v_B}{2\tilde{v}_B}\left\{\Theta_{p,A} + \Theta_{p,B} - 4\pi\Delta\tilde{v}t_p + 2\pi\Delta\tilde{v}T_{link} - 2\pi k_p\right\} \quad (22)$$

after dropping the MF subscripts, which is exactly Eq. (5) of the main text. The second line of Eq. (21) lists the lowest-order errors due to the varying time-of-flight (first two terms) or strong phase drift between cavities. This last term will be insignificant for optical clocks and is even negligible here for the cavity-stabilized lasers. As noted in the text, the first term, $4\pi\tilde{v}_B\Delta t_p\dot{T}_{link}$, dominates the uncertainty for a turbulent link. It contributes a phase error to $\delta\varphi$ of $2\pi v_B\Delta t_p\dot{T}_{link}(t_p)$. If we scale this to units of time, this is an error in the relative timing between the sites of $\Delta t_p\dot{T}_{link}(t_p)$. To convert to a statistical description, the variations in $\dot{T}_{link}(t)$ are characterized by the power spectral density (PSD) $S_{\dot{T}_{link}}(f) = (2\pi f)^2 S_{T_{link}}(f)$, where $S_{T_{link}}(f)$ is the PSD for the atmospheric piston noise in units of time, i.e. of $T_{link}(t)$. As noted in the main text, we approximate this PSD as $S_{Tlink}(f) \approx af^2$. Combining these terms and following the conventional definitions related to Allan deviations, we find an extra phase/timing noise in our clock comparison with a PSD of $S_x = \Delta t_p^2(2\pi f)^2 af^{-2}$ or a corresponding additional fractional frequency fluctuation of $S_y = (2\pi f)^2 S_x = \Delta t_p^2(2\pi f)^4 af^{-2} = (t_{p,B} - t_{p,A})^2 4\pi^4 af^2$. The corresponding modified Allan deviation is

$$\sigma_{MA}^2(t_{avg}) = \frac{3\pi^2 a}{2}(t_{p,B} - t_{p,A})^2 t_{avg}^{-3} \quad (23)$$

as a function of averaging time $t_{avg}$ or Eq. (6) of the main text. We note that with careful time synchronization and simultaneous measurements of $T_{link}(t)$ through the pulse envelope timing, it might be possible, in principle, to include this effect in the analysis and operate at even lower instabilities. However, any experimental demonstration that this is possible can wait until optical clocks have reached even lower levels of instability than their current values.

2. <u>Fade Statistics</u>

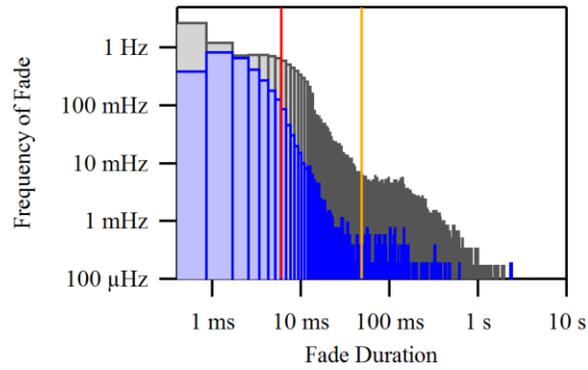

Supplemental Figure 1: Fade statistics for 1% fading of Fig. 2 (blue) and a second run with 26% fading (gray) compared to the mutual coherence times $t_{coh}^{0.12\ rad}$ ~ 7 ms (red line) and $t_{coh}^{1\ rad}$ ~50 ms (orange line). The number of fades of duration greater than $t_{coh}^{0.12\ rad}$ are 1,357 and 28,508 for 1% and 26%-fade runs, respectively.

## 3. Supplemental Material: Kalman-filter phase extraction

Here we will describe the Kalman-filter-based extraction of the carrier phase. The notation here follows the main text except we replace $t_p \to p$ since the digital processing essentially relies on sample number. This extraction takes advantage of all of the measured data, i.e. the time series of the observed unscaled wrapped phase, $\delta\varphi_{obs} = \frac{\tilde{v}_B}{v_B}\delta\varphi(p) + \pi k_p$ (see Eq. (5) in the main text), the time series of the envelope arrival times, $\delta\tau(p)$, the time series of the received power, $I_{rec}(p)$, the dependence of the measured quantities on the received power, $\sigma_{\varphi,obs}(I_{rec})$ and $\sigma_{\tau,obs}(I_{rec})$, and the expected random-walk frequency noise between the lasers from $S_\varphi(f)$. The complicated Kalman-filter-based phase extraction is necessary due to the random distribution and random duration of fades as well as the varying level of measurement noise on the data.

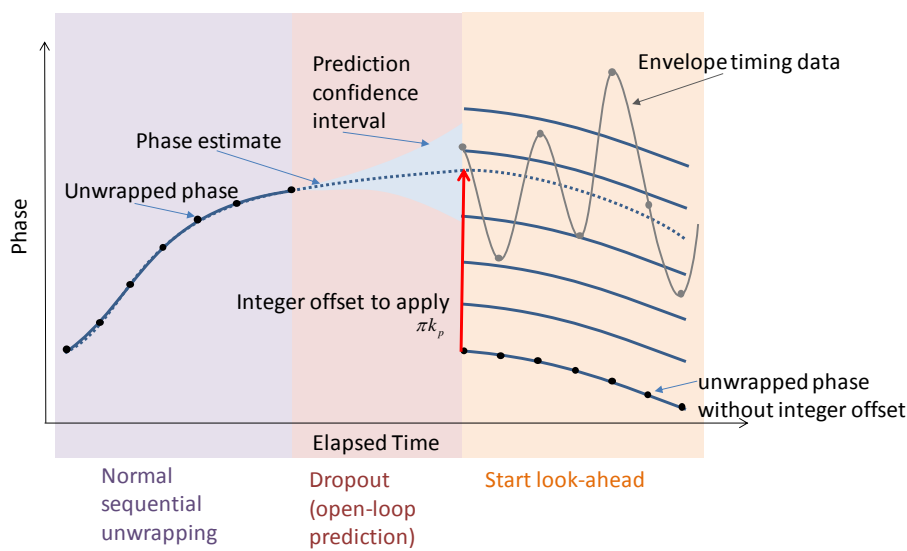

Supplemental Figure 2. Cartoon illustration two modes of extracting the phase, "normal operation" and "look ahead operation". Under normal operation, the phase is sequentially unwrapped using the main Kalman filter. When the prediction confidence interval exceeds a threshold, the system transitions to look-ahead operation to use

up to ~250 ms of data (established empirically) to determine the correct integer offset to unwrap by before returning to normal operation for the next measurement.

The process of extracting the phase has two modes, "normal operation" and "look-ahead operation", depending on the uncertainty of the phase estimate from the Kalman filter. Normal operation is a sequential unwrap of the phase using the main Kalman filter's prediction and is illustrated in Supplemental Fig. 3. (The basic Kalman equations are given below.) In this mode, we first generate a prediction for the unwrapped phase. We compare this predicted value to the observed wrapped phase $\delta\varphi_{obs}(p)$ to resolve the integer, $k_p$. In addition to the prediction, the Kalman filter generates the uncertainty associated with the prediction, $\sigma_{\varphi,p}$. Only if this uncertainty is below a threshold do we consider this $k_p$ to be valid. We use a conservative threshold of 0.12 radians, in order to avoid any integer errors (i.e. phase slips) in $k_p$ since there are ~ 8 million measurements per hour, and to allow for modelling uncertainties. Based on this threshold, we update the main Kalman filter using one of three possibilities: (1) the predicted uncertainty is below threshold, so $k_p$ is valid, and we can proceed to calculate $\delta\varphi(p)$, (2) $\delta\varphi_{obs}(p)$ was unavailable due to a fade, (3) the predicted uncertainty is above threshold, $k_p$ is not necessarily valid, and we switch to the "look-ahead mode".

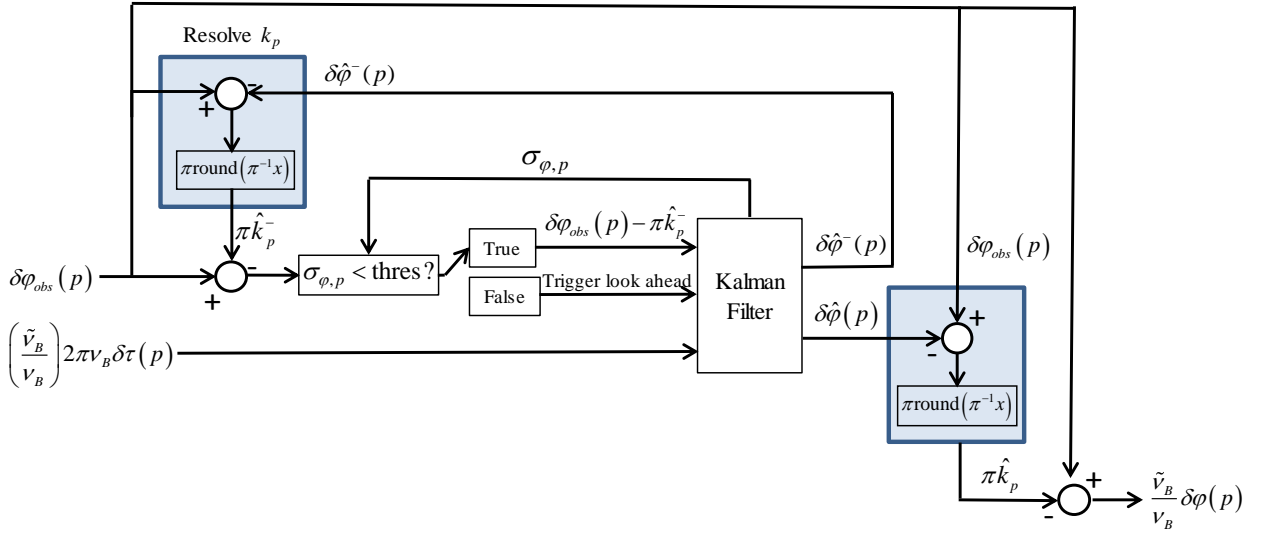

Supplemental Figure 3. Normal operation of Kalman-filter-based phase extraction. The measurement inputs consist of the wrapped phase data $\delta\varphi_{obs}(p) = \frac{\tilde{v}_B}{v_B}\delta\varphi(p) + \pi k_p$ and the envelope timing data scaled appropriately, $\left(\frac{\tilde{v}_B}{v_B}\right) 2\pi v_B \delta\tau(p)$. The *a priori* estimate of the phase, $\delta\hat{\varphi}^-(p)$ is used to generate an initial estimate of the integer offset, $\hat{k}_p^-$. If the uncertainty in the estimate, $\sigma_{\varphi,p}$ is below the threshold, the Kalman filter generates a final estimate of the phase, $\delta\hat{\varphi}(p)$, which is then used to generate a final estimate of $\hat{k}_p$ and thus the unwrapped phase, $\frac{\tilde{v}_B}{v_B}\delta\varphi(p)$. (Note that the phase is evaluated at $\tilde{v}_B$, hence the scaling.) Kalman filter inputs not shown: $\sigma_{\varphi,obs}(I_{rec})$, $\sigma_{\tau,obs}(I_{rec})$, $I_{rec}(p)$ and $S_\varphi(f)$. thres: threshold value for $\sigma_{\varphi,p}$

The second mode, "look-ahead operation", is illustrated in Supplemental Figure 4. It is a more complex operation, as we must find $k_p$ based on the phase data and the envelope timing. It cannot rely on a single measurement as the uncertainty in a single envelope-timing measurement is at best +/- 3 fs. However, simple averaging of the envelope timing is not robust because of the very real possibility of additional fades over the averaging window and because the actual clock noise prevents satisfactory averaging. Instead, two Kalman filters are used. One performs a sequential unwrap of the phase as in normal operation starting at the initial data point $p$ after the fade and letting $k_p = 0$. The difference between this unwrapped phase with $k_p = 0$ and the scaled envelope timing data should be exactly $\pi k_p$ except for noise and any offset between the envelope timing data and the phase data. This offset is transceiver dependent, calibrated in an initial measurement, and applied to the envelope timing data after scaling. The difference between the unwrapped phase and the envelope data is then input to a second Kalman filter along with the measurement uncertainties. After ~250-ms (600 data points) if the uncertainty is sufficiently small, then the estimated $k_p$ is considered valid. It is applied to the already partially unwrapped phase data and the algorithm returns to "normal operation". If there is a fade within the 250-ms but the uncertainty is already sufficiently low, then again $k_p$ is applied to the partially unwrapped phase data and the algorithm returns to "normal operation". If the uncertainty is too high, this measurement is recorded as a fade and the look-ahead operation jumps ahead to unwrap the next set of data again using the look-ahead operation.

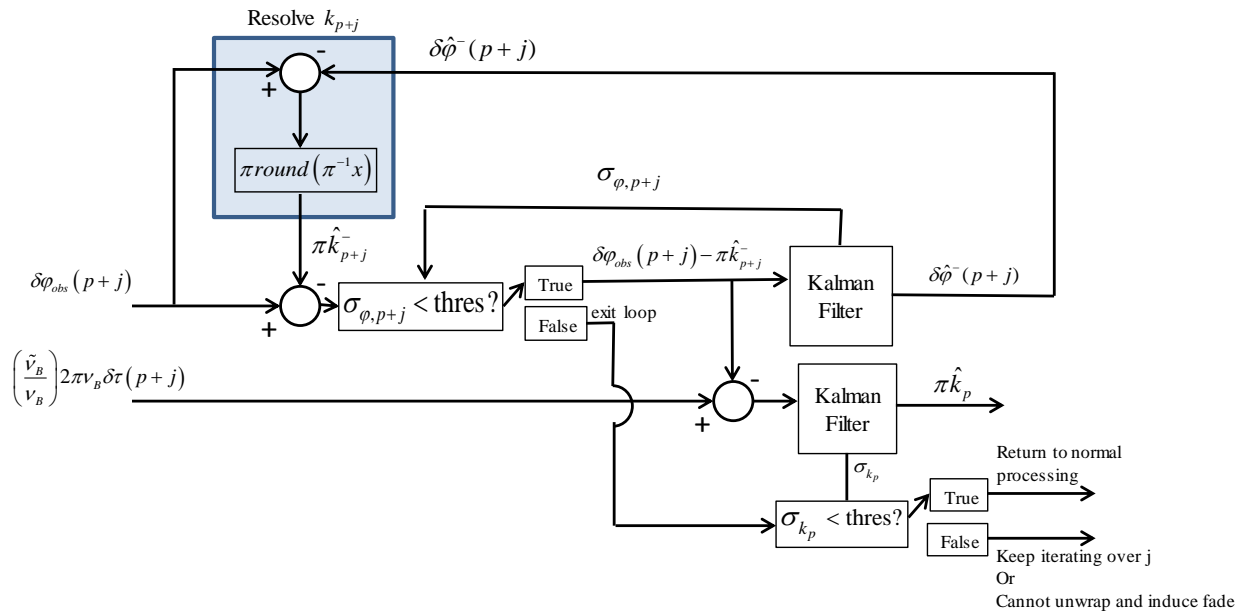

Supplemental Figure 4. Block diagram showing steps for look-ahead operation. Look-ahead operation iterates over index j to determine the correct integer offset to unwrap the p[th] measurement. thres: threshold value for $\sigma_{\varphi,p}$

All of the data presented here were processed post data acquisition and the algorithms were not optimized for speed. However, the method outlined here could be implemented in realtime with a short processing delay for the look ahead operation to provide a real time measure of the clock difference (250 ms delayed) or even to synchronize the phase of the two oscillators by feeding back at low bandwidth.

## 4. Supplemental Material: Kalman filter implementation

We define our state model for the $p^{\text{th}}$ measurement as $\Phi_p = A\Phi_{p-1} + Q_p$ where $\Phi_p = \begin{pmatrix} \delta\varphi_p \\ \delta\dot\varphi_p \end{pmatrix}$, $\delta\varphi_p$ is the phase, $A = \begin{pmatrix} 1 & 2\pi\Delta \\ 0 & 1 \end{pmatrix}$, $1/\Delta = \Delta f_r$ is the sampling rate of the system, and the process noise $Q_p = \begin{pmatrix} 0 & 0 \\ 0 & q_0 2\pi^2 \Delta \end{pmatrix}$ captures the $f^{-4}$ relative oscillator phase noise with $q_0 = 22$ Hz$^2$/Hz to match the measured PSD. Here, we assume that the process noise is constant over the entire measurement duration, i.e. $Q_p = Q_0$. For each measurement point, we generate an *a priori* estimate of the state and of the associated co-variance matrix, $P_p$, via

$$\hat\Phi_p^- = A\hat\Phi_{p-1}$$
$$P_p^- = AP_{p-1}A^T + Q_0 \tag{24}$$

where the ^ symbol represents an estimate of the state and the − symbol indicates an *a priori* value. We can generate an initial estimate of $k_p$ from $\hat\Phi_p^-$. The filter gains, $K_p$ are then computed using the amplitude dependent measurement noise co-variance matrix,

$R_p = \begin{pmatrix} \sigma^2_{\varphi,obs}(I_{rec}) & 0 \\ 0 & \sigma^2_{\tau,obs}(I_{rec}) \end{pmatrix}$, as is the posterior estimate of the state, $\hat\Phi_p$, and the posterior associated co-variance matrix, $P_p$

$$K_p = P_p^- H \left(HP_p^- H^T + R_p\right)^{-1}$$
$$\hat\Phi_p = \hat\Phi_p^- + K_p\left(Z_p - H\hat\Phi_p^-\right) \tag{25}$$
$$P_p = \left(I - K_p H\right) P_p^-$$

where $I$ is the identity matrix, $Z_p$ is a set of measurements, and $H$ is the matrix which relates the underlying state to the measurements, i.e. $H\Phi_p = Z_p$. The posterior estimate of the state, $\hat{\Phi}_p$, is then used to generate the final estimate of $k_p$.

Note that in the presence of a fade, the posterior equations of (0.8) reduce to $\hat{\Phi}_p = \hat{\Phi}_p^-$ and $P_p = P_p^-$. From this, we can immediately see how the co-variance matrix for the uncertainty of the state estimate evolves as the fade duration increases. We can write the recursive expression for fades of increasing length to find $P_{p+n} = A^{n+1} P_{p-1} A^{Tn+1} + \sum_0^n \left( A^n Q_0 A^{Tn} \right)$ where n is the number of samples for the fade. This yields

$$\sigma_{\varphi,p+n}^2 \approx \sigma_{\varphi,p-1}^2 + 2\pi\Delta(n+1)\left(\sigma_{\varphi\dot{\varphi},p-1}^2 + \sigma_{\dot{\varphi}\varphi,p-1}^2\right) + \left[2\pi\Delta(n+1)\right]^2 \sigma_{\dot{\varphi},p-1}^2$$
$$+ \frac{4}{3}\pi^4 q_0 \Delta^3 n(n+1)(1+2n)$$
(26)

where the co-variance matrix $P_{p-1}$ is defined as $P_{p-1} = \begin{pmatrix} \sigma_{\varphi,p-1}^2 & \sigma_{\varphi\dot{\varphi},p-1}^2 \\ \sigma_{\dot{\varphi}\varphi,p-1}^2 & \sigma_{\dot{\varphi},p-1}^2 \end{pmatrix}$. For $\sigma_{\varphi,p-1} \approx 0.05$ rad, $\sigma_{\varphi\dot{\varphi},p-1} = \sigma_{\dot{\varphi}\varphi,p-1} \approx 0.2$ Hz$^{1/2}$, $\sigma_{\dot{\varphi},p-1} = 2$ Hz, $\Delta \approx 0.4$ ms and $q_0 = 22$ Hz$^2$/Hz, this results in coherence times of $t_{coh}^{0.12\,\text{rad}} \sim 6$ ms and $t_{coh}^{1\,\text{rad}} \sim 50$ ms which agree with the values extracted from the measurements.

5. <u>Supplemental Table I</u>

Supplemental Table I. Experimental values. The symbols are defined in the text. The values assume that the comb A repetition rate is exactly 200 MHz (as we assume it defines the timebase). The actual comb A repetition rate compared to a Hydrogen maser was 200.733 MHz. Values reported against the maser timebase, should all be scaled accordingly.

| | |
|---|---|
| $n_A$ | **972920** |
| $\tilde{n}_A$ | 974276 |
| $n_B$ | 972909 |
| $\tilde{n}_B$ | 974264 |
| $f_{r,A}$ | 200 MHz |
| $f_{r,B}$ | 200.002463 MHz |
| $\Delta f_r$ | 2.464 kHz |
| $\nu_A$ | 194.584000 THz |
| $\tilde{\nu}_A$ | 194.855225 THz |
| $\Delta \nu$ | 196.7 MHz |
| $\Delta \tilde{\nu}$ | -1.2 kHz |